\documentclass[12pt]{article}
\baselineskip %
\advance %
\textheight by %
\topskip %
\usepackage{amsfonts,amsthm,amssymb}
\usepackage{amsfonts}
\usepackage{amsmath}
\usepackage{amssymb}

\usepackage{multirow}

\usepackage{verbatim}

\usepackage{dcolumn}
\usepackage{color}
\usepackage{graphicx}
\usepackage{mathrsfs}
\usepackage{graphicx}

\usepackage{graphicx}

\usepackage{graphicx}

\definecolor{DarkBlue}{rgb}{0,0.1,0.7}

\newcommand{\dd}{\mathrm{d}}
\newcommand{\R}{{\mathord{\mathbb R}}}

\newcommand{\N}{{\mathord{\mathbb N}}}
\newcommand{\ee}{\mathrm{e}}

\begin{document}

\title{Dispersionless wave packets in Dirac materials}
\author{{V\'i{}t Jakubsk\'y$^\dagger$, Mat\v ej Tu\v sek$^\ddagger$}\\
{\small 
\textit{$^\dagger$Department of Theoretical Physics,
Nuclear Physics Institute,
25068 \v Re\v z, Czech Republic}}\\
{\small \textit{$^\ddagger$Department of Mathematics, Czech Technical University in Prague,}}\\{\small\textit{Trojanova 13, 120 00 Prague, Czech Republic}}\\
\sl{\small{E-mail: jakub@ujf.cas.cz,  tusekmat@fjfi.cvut.cz
} }}
\date{\today}
\maketitle

\begin{abstract}
We show that a wide class of quantum systems with translational invariance can host dispersionless, soliton-like, wave packets. We focus on the setting where the effective, two-dimensional Hamiltonian acquires the form of the Dirac operator. The proposed framework for construction of the dispersionless wave packets is illustrated on  silicene-like systems with topologically nontrivial effective mass. Our analytical predictions are accompanied by a numerical analysis and possible experimental realizations are discussed.
 
\end{abstract}

\section{Introduction} 
Dispersion of wave packets is usually considered as a hallmark of quantum systems. However, it can be avoided.
The seek for non-dispersive quantum wave packets dates back to Erwin Schr\"odinger. In 1926, he introduced a wave function whose maximum of amplitude followed classical trajectory in the field of harmonic oscillator \cite{Schrodinger}. Later on, a non-dispersive (but not normalizable) wave function was constructed for a free particle system in terms of the Airy functions \cite{berry}. Other milestones were carved by prediction and preparation of non-dispersive wave packets in experiments with the Rydberg atoms. Wave packets orbiting the atom were stabilized by an external periodic perturbation represented by a fine-tuned electromagnetic wave \cite{trojan1}-\cite{trojan3}. An analogous phenomenon was discussed recently in the context of many body systems where the wave packets were composed of quasi-particles \cite{NJP16}.

A surprising variety of condensed matter systems shares one distinctive feature; low-energy quasi-particles behave like massless or massive Dirac fermions. These systems were coined as the Dirac materials in the literature \cite{Wehling}. The graphene, despite being a prominent representative of the family, is not its only member. The Dirac fermions were predicted in the silicene, germanene or dichalcogenides \cite{SiGeTheor1}-\cite{dichalcogenides}. They are present in high-temperature d-wave superconductors \cite{sucond1} or superfluid phases of ${}^3$He \cite{suflu2}.
The Dirac fermions emerge naturally in the low-energy approximation of the tight-binding model  of a generic hexagonal lattice \cite{DiracSym}. This fact can be utilized for preparation of the artificial graphene. It was created by confining ultracold atoms to  hexagonal optical lattices \cite{opt1}-\cite{opt3}, by assembling  carbon mono\-xide molecules into hexagonal lattices on a copper surface \cite{molecular}, by drilling holes into a Plexiglas plate \cite{ac1} or simulated in other experiments with acoustic waves \cite{ac2}. See also a review article \cite{agr1}.  

In this paper, we discuss a class of quantum systems where dispersion of the wave packets can be suppressed due to the symmetries and intrinsic spectral properties of the considered setting. First, we consider an abstract quantum system and specify sufficient conditions for hosting of the non-dispersive wave packets. A physical realization of a system described by the Dirac equation is introduced later on. Both numerical and qualitative analysis are carried on and an experimental realization is discussed. 

\section{Absence of dispersion }
Let us consider a two-dimensional quantum system that possesses translational invariance along $y$-axis, i.e., its Hamiltonian $H(x,y)$ commutes with the generator of translations $\hat{k}_y=-i\hbar\partial_y $,
\begin{equation}\label{eq:comm}
 [H(x,y),\hat{k}_y]=0.
\end{equation}
It is convenient to rewrite the wave functions in terms of the  generalized eigenstates of $\hat{k}_y$. Using the partial Fourier transform $\mathscr{F}_{y\to k}$ (and its inverse), we obtain
\begin{align}\label{fpsi}
\psi(x,y)=\mathscr{F}^{-1}_{y\to k}\psi(x,k)=(2\pi\hbar)^{-1/2}\int_\R e^{\frac{i}{\hbar}ky}\psi(x,k)\dd k,\\
\psi(x,k)=\mathscr{F}_{y\to k}\psi(x,y)=(2\pi\hbar)^{-1/2}\int_\R e^{-\frac{i}{\hbar}ky}\psi(x,y)\dd y.\nonumber
\end{align}
The multiplicative factor $(2\pi\hbar)^{-1/2}$ guarantees unitarity of the mapping.
The formula (\ref{fpsi}) can be understood as a generalization of the partial wave expansion for the case where the eigenvalues of $\hat{k}_y$ are not quantized. The action of the Hamiltonian can be written as
\begin{equation}\label{eq:DI_decomp}
  H(x,y)\psi(x,y)=(2\pi\hbar)^{-1/2}\int_\R e^{\frac{i}{\hbar}ky} H(x,k)\psi(x,k)\dd k.
\end{equation}
Here, $H(x,k)= \mathscr{F}_{y\to k}H(x,y)\mathscr{F}_{y\to k}^{-1}$ should be understood as an operator acting in the $x$-variable only. It will be referred to  as the \textit{fiber Hamiltonian} later in the text. The formula (\ref{eq:DI_decomp}) says that in the partial momentum representation, our Hamiltonian decomposes into the so-called direct integral of the fiber operators $H(x,k)$, see e.g.  \cite{RS4}.

Now, let us suppose that $H(x,k)$ has a  non-empty set of discrete eigenvalues for $k$ from an open interval $J\subset\R$. We denote them $E_n(k)$ where the positive integer $n$ labels the energy bands. The associated normalized bound states $F_n(x,k)$ satisfy 
\begin{equation}\label{H(x,k)E}
(H(x,k)-E_n(k))F_n(x,k)=0,\quad k\in J.
\end{equation} 
If we suppose $H(x,k)$ to be analytic in $k$ then the functions $E_n(k)$ are analytic in $k$ as well \cite{kato}.

We can construct a wave function $\Psi_n(x,y)$ using the eigenstates corresponding to a single energy band $E_n(k)$, 
\begin{equation}\label{Psi_n}
 \Psi_n(x,y)=(2\pi\hbar)^{-1/2}\int_{I_n}e^{\frac{i}{\hbar}ky}\beta_n(k)F_n(x,k)\dd k.
\end{equation}
The coefficient function $\beta_n(k)$ is supported by $I_n\subset J$, i.e., $\beta_n(k)=0$ for all $k\notin I_n$.
We require $\Psi_{n}$ to be normalized. This is guaranteed as long as $\int_{I}|\beta_{n}(k)|^2dk=1$. 

In general, the wave packet (\ref{Psi_n}) can get dispersed along the $y$-axis. However, by construction, $\Psi_n$ has a remarkable transverse stability,  it is dispersionless along the $x$-axis. This transverse stability does not stem from a fine interference of non-localized waves but from the fact that each $F_n(x,k)$ is an eigenfunction of $H(x,k)$. Indeed, we have
\begin{multline}
 \int_\R|\ee^{-\frac{i}{\hbar}tH(x,y)}\Psi_n(x,y)|^2\dd y=\int_\R|\ee^{-\frac{i}{\hbar}tH(x,k)}\Psi_n(x,k)|^2\dd k=\int_\R|\ee^{-\frac{i}{\hbar}tE_n(k)}\Psi_n(x,k)|^2\dd k\\
 =\int_\R|\Psi_n(x,y)|^2\dd y.
\end{multline}
In the first equality we used unitarity of $\mathscr{F}_{y\to k}$. Therefore, the probability density of finding the particle at distance $x$ from the $y$-axis does not change with time.

Dispersion of the  wave packets (\ref{Psi_n}) can be completely suppressed provided that $E_n(k)$ is linear on the interval $I_n\subset J$,
\begin{equation}\label{eq:lin_ev}
E_n(k)=e_n+v_n k,
\quad k\in I_n\subset J.
\end{equation}
Then the wave packet $\Psi_n$ has a \textit{soliton-like} behavior; it evolves with a uniform speed without any dispersion, 
\begin{equation}\label{nd1}
e^{-\frac{i}{\hbar}tH(x,y)}\Psi_{n}(x,y)=c_n(t)\Psi_{n}(x,y- v_n t),\quad |c_n(t)|=1.
\end{equation} 
Indeed, we have
\begin{eqnarray}
 e^{-\frac{i}{\hbar}tH(x,y)}\Psi_n(x,y)
 &=&(2\pi\hbar)^{-1/2}\int_{I_n}e^{\frac{i}{\hbar}ky}e^{-\frac{i}{\hbar}tH(x,k)}(\beta_n(k)F_n(x,k))\dd k\nonumber\\
 &=&(2\pi\hbar)^{-1/2}\int_{I_n}e^{\frac{i}{\hbar}ky}\beta_n(k)e^{-\frac{i}{\hbar}tH(x,k)}F_n(x,k)\dd k\nonumber\\
 &=&e^{-\frac{i}{\hbar}te_n}(2\pi\hbar)^{-1/2}\int_{I_n}e^{\frac{i}{\hbar}k(y-v_n t)}\beta_n(k)F_n(x,k)\dd k\nonumber\\
&=&e^{-\frac{i}{\hbar}te_n}\Psi_n(x,y-v_n t).
\end{eqnarray}
In the first equality we view $H(x,k)$ as an operator acting in both variables. It obeys $[H(x,k),k]=0$, due to (1). This implies the second equality where $H(x,k)$ is already the fiber operator.
Hence, the wave packet  moves uniformly with the speed $v_n$ that corresponds to the slope of the energy band $E_n(k)$, $v_n=E_n'(k)$,  $k\in I_n$. When $v_n=0$, the wave packet $\Psi_n$ neither moves nor dissipates, it represents a stationary state of $H(x,y)$ with energy $e_{n}$. 

The existence of dispersionless wave packets (\ref{Psi_n}) is rather independent of the explicit form  of $H(x,y)$. The only assumptions on $H(x,y)$ are just the commutation relation (\ref{eq:comm}), which implies the decomposition \eqref{eq:DI_decomp},  and the favorable spectral properties described by (\ref{H(x,k)E}). Under these assumptions, $\Psi_n$ does not disperse along $x$-axis as has been already discussed above. If, in addition,  (\ref{eq:lin_ev}) holds then $\Psi_n$ is dispersionless \eqref{nd1} in the both directions and moves along the $y$-axis with the uniform velocity $v_n$.

It is straightforward to generalize the framework to a multidimensional Hamiltonian. If $\vec{x}\in\R^l$ and $\vec{k}\in\R^m$ are multidimensional vectors of not necessarily the same dimension, $[H(\vec{x},\vec{k}),\vec{k}]=0$, and  $E_n=e_n+\vec{v}_n\cdot \vec{k}$, where $\vec{k}\in I_n\subset \R^m$, then the wave packet $\Psi_n$ will be dispersionless and moving in $(\vec{x},\vec{y})$--space with the velocity vector $(0,\vec{v}_n)$.

Though we are not aware of any non-relativistic system where (\ref{eq:lin_ev}) would take place, there is an elegant way how to guarantee existence of a \textit{globally} linear energy band in the Dirac materials under some specific conditions. We shall discuss it in Section \ref{model}.

\section{Perturbing the dispersionless wave packets}
In this section, we consider perturbations of a system hosting dispersionless wave packets. We consider two scenarios. Firstly, the perturbation is taken into the account implicitly as the cause for the breakdown of the linear dispersion relation (\ref{eq:lin_ev}). In the second case, we study the effect of a weak perturbation $V(x,y)$ on the propagation of the wave packets. In both cases, we compare the exact time evolution of the wave packet $\Psi_n$ affected by the perturbation with the wave packet $\Psi_n$ evolving in a soliton-like manner with a uniform speed $v$. Our goal is to specify the transition amplitude
\begin{eqnarray}\label{A_defbasic}
&&A(t)=\langle \Psi_n(x,y-vt),e^{-\frac{i}{\hbar}tH(x,y)}\Psi_n(x,y)\rangle,
\end{eqnarray}
and to determine the optimal value of $v$ that maximizes the transition probability $|A(t)|^2$. This value would  approximate  the speed of propagation of the dispersing wave packet.

Let us suppose that $E_n(k)$ is not li\-near.  Then the wave packet \eqref{Psi_n} can still disperse very slowly along the $y$-axis provided that $E_n(k)$  is well approximated by a linear function on the interval $I_n$. 
We denote the deviation of the energy band $E_n(k)$ from a linear function by
\begin{equation}\label{B}
B(k)=E_n(k)-(e+vk),\quad k\in I_n,
\end{equation}
where $e$ and $v$ are some real constants. Then a straightforward calculation reveals that 
\begin{equation}\label{A_def}
|A(t)|^2=\int_{I_n\times I_n}dkds |\beta_n(k)|^2|\beta_n(s)|^2\cos \left((B(k)-B(s))\frac{t}{\hbar}\right).
\end{equation}
Notice that the integrand does not depend on $e$.  A short analysis of the stationary equation $\partial_v |A(t)|^2=0$ shows that the maximizing value of $v$ depends on $\beta$ and $t$ and cannot be, in general, written in a closed form. For small values of $t$, it is given approximately by $\frac{E_n(k)-E_n(s)}{k-s}$ averaged over $I_n\times I_n$ with the weight $|\beta_n(k)|^2|\beta_n(s)|^2(k-s)^2$.

Let us find an estimate for $v$ that would be reasonably simple and accurate, yet possibly not maximizing $|A(t)|^2$. To this purpose we start with the following estimates
\begin{eqnarray}
&&|A(t)|^2\geq\inf_{(k,s)\in I_n\times I_n}\cos\left((B(k)-B(s))\frac{t}{\hbar}\right)\nonumber\\
 &&\geq 1-\frac{t^2}{2\hbar^2}\sup_{(k,s)\in I_n\times I_n}(B(k)-B(s))^2
 \label{a2}\\&&\geq 1-\frac{2t^2}{\hbar^2}\sup_{k\in I_n}|B(k)|^2\label{a3}.
\end{eqnarray}
Now we find the value of $v$ that maximizes the lower bound.  This can be done in a straightforward manner if $I_n=(a,b)$ is chosen in a way that for all $k\in I_n$, the graph of $E_n(k)$  lies under or above the line segment connecting its endpoints.  Then \eqref{a3} coincides with \eqref{a2} and the expression can be optimized by putting $v$ equal to the averaged \textit{group velocity}, i.e.,
\begin{equation}\label{eq:prop_vel}
v=\frac{\int_{I_n}E'_n(k)dk}{b-a}=\frac{E_n(b)-E_n(a)}{b-a}.
\end{equation}
The value of $e$ is then to be fixed so that $e+vk$ goes in the middle of the narrowest strip containing $E_n(k)$, i.e., $\sup_{k\in I_n}(E_n(k)-vk-e)=-\inf_{k\in I_n}(E_n(k)-vk-e)$. The approximation of $v$ by the averaged group velocity is in a very good agreement with our numerical results depicted in Fig. \ref{fig:numerics} (a)--c)) for three different packets that disperse slowly along the $y$-axis. There, the small white cross moves uniformly with the velocity (\ref{eq:prop_vel}) and approximates very well the exact motion of the center of mass of the wave packet. We also plotted there the transition probability $|A(t)|^2$  ( with $v$ given by \eqref{eq:prop_vel}) for the three definite choices of $\Psi_n$.

Now, let us turn our attention to the second scenario in which the Hamiltonian decomposes as follows
\begin{equation}\label{HV}
 H(x,y)=H_0(x,y)+V(x,y),
\end{equation}
where the system described by $H_0$ can host dispersionless wave packets and $V$ is a bounded symmetric operator, e.g., the multiplication operator by a bounded real function, that plays role of a perturbation. 
In order to describe the time evolution of the system with the Hamiltonian $H$, it is convenient to employ the Dyson expansion \cite{BlExHa}, 
\begin{multline*}
 \psi[t]=\ee^{-\frac{i}{\hbar}t H}\psi\\=\ee^{-\frac{i}{\hbar}tH_0}\psi+\sum_{j=1}^{+\infty}(-i)^j\int_0^t\dd t_1\int_0^{t_1}\dd t_2\ldots\int_0^{t_{j-1}}\dd t_j\ee^{-\frac{i}{\hbar}(t-t_1)H_0}V\ee^{-\frac{i}{\hbar}(t_1-t_2)H_0}V\ldots\\
 \ee^{-\frac{i}{\hbar}(t_{j-1}-t_j)H_0}V\ee^{-\frac{i}{\hbar}t_j H_0}\psi,
\end{multline*}
where $\psi\equiv\psi[0]$ is a normalized state at $t=0$ and $\psi[t]$ is the state at time $t$. 
The transition amplitude is given in terms of the expansion coefficients $\alpha_j$,
	\begin{equation}\label{eq:transition}
	\langle\ee^{-\frac{i}{\hbar}tH_0}\psi,\ee^{-\frac{i}{\hbar}tH}\psi\rangle=1+\sum_{j=1}^{+\infty}(-i)^j \alpha_j,
	\end{equation}
	where 
	\begin{equation*}
	\alpha_j=\int_0^t\dd t_1\int_0^{t_1}\dd t_2\ldots\int_0^{t_{j-1}}\dd t_j\langle\psi,\ee^{\frac{i}{\hbar}t_1H_0}V\ee^{-\frac{i}{\hbar}(t_1-t_2)H_0}V\ldots
	\ee^{-\frac{i}{\hbar}(t_{j-1}-t_j)H_0}V\ee^{-\frac{i}{\hbar}t_j H_0}\psi\rangle.
	\end{equation*}
In Appendix, we present a rather technical computation of the expansion for the transition probability $|\langle\ee^{-\frac{i}{\hbar}tH_0}\psi,\ee^{-\frac{i}{\hbar}tH}\psi\rangle|^2$.

Let us take a dispersionless state $\Psi_n$ of the unperturbed system for $\psi$, i.e., $(\ee^{-\frac{i}{\hbar}tH_0}\Psi_n)(x,y)=\ee^{-\frac{i}{\hbar}e_n t}\Psi_n(x,y-v_nt)$.  For $\|V\|t\ll 1$, the transition probability is well approximated by \eqref{eq:asy}, i.e.,
\begin{equation}\label{eq:asy_text}
 |A(t)|^2=1-\tilde\alpha_2(t)+\alpha_1(t)^2+\mathcal{O}(t^3),
\end{equation}
where $\alpha_1(t)$ and $\tilde\alpha_2(t)$ are given by 
\begin{align}
 &\alpha_1(t)=\int_0^t\dd t_1\langle\Psi_n(x,y-v_nt_1),V(x,y)\Psi_n(x,y-v_nt_1)\rangle\label{eq:Dyson_1}\\
 &\tilde\alpha_2(t)=\int_0^t\dd t_1\int_0^t\dd t_2\,\ee^{\frac{i}{\hbar}e_n(t_1-t_2)}\langle \ee^{\frac{i}{\hbar}t_1 H_0}V(x,y)\Psi_n(x,y-v_nt_1),\ee^{\frac{i}{\hbar}t_2 H_0}V(x,y)\Psi_n(x,y-v_nt_2)\rangle\label{eq:Dyson_2}.
\end{align}

 The coefficients in the expansion  take particularly simple form, if $V(x,y)$ is just an electrostatic field that is invariant with respect to the translations along the $x$-axis, i.e.,  $V(x,y)$ acts like  $V=W(y)\mathbf{1}$. In that case, we have
\begin{equation}\label{Aj}
\alpha_j=\int_0^t\dd t_1\int_0^{t_1}\dd t_2\ldots\int_0^{t_{j-1}}\dd t_j\langle\Psi_n(x,y),W(y+v_nt_1)W(y+v_nt_2)\ldots W(y+v_nt_j)\Psi_n(x,y)\rangle,
\end{equation}
and, in particular,
\begin{align}
 &\alpha_1(t)=\int_0^t\dd t_1\langle\Psi_n(x,y),W(y+v_nt_1)\Psi_n(x,y)\rangle=\langle\Psi_n(x,y),\tilde W(y;t)\Psi_n(x,y)\rangle \label{Dyson_1_spec}\\
 &\tilde\alpha_2(t)\begin{aligned}[t]&=\int_0^t\dd t_1\int_0^t\dd t_2\langle\Psi_n(x,y),W(y+v_nt_1)W(y+v_nt_2)\Psi_n(x,y)\rangle\\
 &=\langle\Psi_n(x,y),\tilde W(y;t)^2\Psi_n(x,y)\rangle,
\end{aligned} \label{Dyson_2_spec}
\end{align}
where $\tilde W(y;t):=\int_0^t\dd t_1 W(y+v_nt_1)$.

Evaluation of $\alpha_1$ and $\tilde\alpha_2$ is impossible without  the explicit knowledge of $V(x,y)$. 
However, by construction, the wave packet $\Psi_n(x,y)$ decays rapidly along the $x$-axis and it is dispersionless in the transverse direction. When $V=V(x,y)\mathbf{1}$ is localized far enough from the $y$-axis then the both  $\alpha_1$ and $\tilde\alpha_2$ get strongly suppressed since the overlap of $V(x,y)$ and $\Psi_n(x,y-v_nt)$, that is important for a contribution to the  integrals \eqref{eq:Dyson_1} and \eqref{eq:Dyson_2}, tends to zero for any $t$.

\section{Model for dispersionless wave packets\label{model}}
In the graphene or silicene (and other systems sharing qualitatively the same tight-binding Hamiltonian), the valence and conduction bands touch in discrete points that correspond to the corners of the first Brillouin zone. The dispersion relation is linear there, justifying to call them the Dirac points. In their vicinity, behavior of quasi-particles is described by two Dirac equations where the Dirac operator can be represented in terms of $2\times2$ matrices \cite{semenoff}. The spinor structure of wave functions reflects presence of two triangular sublattices $A$ and $B$ in the lattice. The two equations, governing dynamics at the vicinity of  two inequivalent Dirac points $K$ and $K'$, are uncoupled provided that interactions do not cause inter-valley scattering. We suppose this to be the case.

We fix the  Hamiltonian in the following form
\begin{equation}\label{hm}
 H(x,y)= v_F\tau_3\otimes\left(-i\hbar\sigma_1\partial_x-i\hbar \sigma_2\partial_{y}+\frac{\gamma_0}{v_F}m(x)\sigma_3\right),
\end{equation}
whose fiber operator reads
\begin{equation*}
 H(x,k)= v_F\tau_3\otimes\left(-i\hbar\sigma_1\partial_x+k\sigma_2+\frac{\gamma_0}{v_F}m(x)\sigma_3\right).
\end{equation*}
Here, $\tau_a$ and $\sigma_a$ are the Pauli matrices acting on the valley and sublattice degree of freedom, respectively.  The bispinors have the following structure, $\Psi=(\psi_{K,A},\psi_{K,B},\psi_{K',B},\psi_{K',A})^T$, where the first index denotes the valley whereas the second one stays for the sublattice. 
The term $m(x)$ represents an effective mass, $v_F=\frac{3}{2\hbar}a_{cc}\gamma_0$ is the Fermi velocity,  $\gamma_0$ is the hopping energy and $a_{cc}$ is the distance between two sites in the hexagonal lattice. In the graphene, $v_F$ equals approximately $1/300$ of the speed of light, $\gamma_0\sim 2.3\,$\mbox{eV} and $a_{cc}\sim 0.142\times 10^{-9}\,$m. 

We fix the mass term $m(x)$ so that it is a bounded (but not necessarily continuous) function and acquires a topologically nontrivial form,
\begin{equation}\label{mass}
 \lim_{x\rightarrow\pm\infty}m(x)=m_{\pm},\quad  m_-<0,\quad m_+>0.
\end{equation}
Then the fiber Hamiltonian $H(x,k)$ possesses two nodeless bound states that are localized at the domain wall where the mass changes sign \cite{Semenoff}. Their explicit (not normalized) form is
\begin{eqnarray}
&&F_+(x)=\left(1,i,0,0\right)^Te^{-\frac{\gamma_0}{\hbar v_F}\int_0^xm(s)ds},\nonumber\\
&&F_-(x)= \tau_1\otimes\sigma_2\, F_{+}(x)=\left(0,0,1,i\right)^Te^{-\frac{\gamma_0}{\hbar v_F}\int_0^xm(s)ds}.\label{FF}\nonumber
\end{eqnarray}
They satisfy
\begin{equation}\label{ia}
 H(x,k)F_{\pm}(x)=\pm v_F k F_{\pm}(x)
\end{equation}
and can be utilized to prepare spatially well localized wave packets $\Psi_{\pm}$. The spatial localization comes at a price of an extended interval where the coefficient function $\beta_{\pm}(k)$ in (\ref{Psi_n}) is nonvanishing. However, it does not compromise  the relation (\ref{eq:lin_ev}), as (\ref{ia}) is valid for any $k\in\R$. 
In coordinate representation, the non-dispersive wave packets  can be  written in a particularly simple form, 
\begin{equation}\label{nds}
 \Psi_{\pm}(x,y)=G_{\pm}(y)F_{\pm}(x), 
\end{equation}
where $G_{\pm}(y)$ are arbitrary square integrable scalar functions.

The Hamiltonian (\ref{hm}) commutes with the operator  $S=R_y\tau_1\otimes\sigma_2$  where $R_y$ is the reflection operator, i.e., $(R_y\psi)(x,y)=\psi(x,-y)$. The symmetry $S$ tells us that, for any dispersionless wave packet $\Psi_n$, there exists a counter-propagating dispersionless wave packet $\Phi_n=S\Psi_n$ which has the same expectation value of energy,
\begin{equation}\label{nd2}
e^{-\frac{i}{\hbar}tH(x,y)}\Phi_{n}(x,y)=c_n(t)\,\Phi_{n}(x,y+ v_n t).
\end{equation}
We define the valley-projectors $\Gamma^{\pm}=\frac{1}{2}(\mathbf{1}\pm\tau_3\otimes \mathbf{1})$. When $\Psi_n$ consists of states from the $K$-valley, $\Gamma^+\Psi_n=\Psi_n$, then $\Phi_n$ is $K'$-valley polarized, $\Gamma^-\Phi_n=\Phi_n$. As there is no interaction between the valleys ($H$ is a block-diagonal operator), the two valley-polarized packets can propagate simultaneously without scattering \footnote{The Hamiltonian anticommutes with $\tau_a\otimes\mathbf{1}$, $a=1,2$. Hence, there exists a non-dispersive state $\Phi_{n}=\tau_1\otimes\mathbf{1}\Psi_n$ that also satisfies (\ref{nd2}) with $c_n(t)\rightarrow \overline{c_n(t)}$.}. The system represents a loss-less communication channel for possible valleytronics devices. 

\begin{figure}[h!]
	\centering
	\includegraphics[scale=.6]{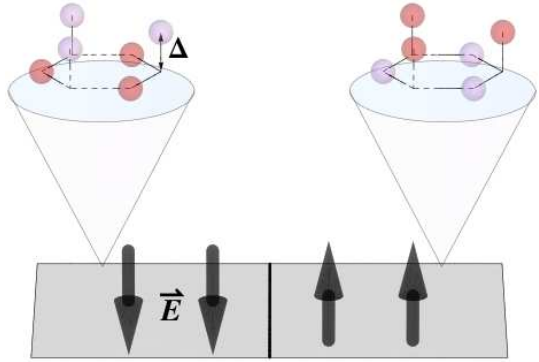}
	\caption{(color online) Topologically nontrivial mass for a buckled hexagonal lattice. The hexagonal lattice (gray plane) has buckling parameter $\Delta$ (in the magnified buckled hexagons).  Black arrows denote direction of the intensity of electric  field $\vec{E}$ and the black line between the arrows marks the threshold where the intensity changes the sign. Sublattice symmetry is broken as each of the two  triangular sublattices feels different electrostatic potential (its value on the sublattices is illustrated by the red and white color of the atoms). \label{zeromodesgap}}
	\label{polyhedra}
\end{figure}

The effective mass (\ref{mass}) can be associated with a specific breakdown of the sublattice symmetry of the hexagonal lattice. It can be realized experimentally, e.g., in the silicene. There, the breakdown can be caused by an electric field perpendicular to the crystal plane \cite{EffMSilicene} that changes orientation when passing from the region with $x<0$ to the region with $x>0$, see Fig.\ref{polyhedra} for illustration.  
Recent elaboration of a field-effective transistor based on the sili\-cene \cite{FET} suggests that experimental realization of such setup should be feasible. The artificial graphene could be another host for preparation of the systems with (\ref{hm}) and (\ref{mass}), the molecular graphene \cite{molecular} with its tunable properties in particular. Let us also mention that the effective mass (\ref{mass}) appears on the interfaces of the quantum anomalous Hall and quantum valley Hall insulators discussed 
recently in \cite{PRB92}. 

\subsection*{Example}
 We start with the substitution $x\mapsto \frac{2}{3a_{cc}} x$, $y\mapsto \frac{2}{3a_{cc}} y$, $k\mapsto\frac{3a_{cc}}{2\hbar}k$, $E\mapsto E/\gamma_0 $, and $t\mapsto\frac{\gamma_0}{\hbar}t$. In these units, $v_F\hbar=1$ and $v$ is measured in multiples of $v_F$. Next we fix $m(x)=\alpha \omega \tanh (\alpha x)$, where $\alpha\in\R$ and $\omega>0$. We focus on the analysis of the $K$-valley. The behavior of the system in the $K'$-valley can be acquired directly as the corresponding energy operators differ just in sign, see (\ref{hm}). For convenience, we employ a unitary transformation such that  $\sigma_1\mapsto\sigma_1$, $\sigma_2\mapsto \sigma_3$, and $\sigma_3\mapsto -\sigma_2$, and denote transformed quantities by tilde. The effective Hamiltonian in the $K$-valley then reads
\begin{equation}\label{Hexample}
\tilde{H}_K(x,k)=-i\sigma_1\partial_x-\omega\alpha \tanh (\alpha x)\sigma_2+k\sigma_3.
\end{equation}

The fiber Hamiltonian $\tilde{H}_K(x,k)$ has both discrete ener\-gies and the continuous spectrum. We label the bound states  by $n\in\{0,1,\dots,\lfloor\omega\rfloor\}$, where $\lfloor.\rfloor$ stands for the integer part. For $n\neq 0$, the (non-normalized) bound states $\tilde{F}_{n}^{\pm}(x,k)$ of $\tilde{H}_K(x,k)$ satisfy 
\begin{equation}\nonumber
 \tilde{H}_K(x,k)\tilde{F}_n^{\pm}(x,k)=\pm E_n(k)\tilde{F}_n^{\pm}(x,k),
\end{equation}
where $E_n(k)=\sqrt{n(-n+2\omega)\alpha^2+k^2}$.
They can be written as  
\begin{equation*}
 \tilde{F}_{n}^{\pm}(x,k)=\begin{pmatrix}1&0\\0&\epsilon^{\pm}(k,n)\end{pmatrix}\left(\mathbf{1}+\frac{\tilde{H}_K(x,0)}{E_n(0)^2}\right)\left(\begin{array}{c}f_{n}(x)\\0\end{array}\right)
\end{equation*}
\cite{isomorph}, where we denoted $\epsilon^{\pm}(k,n)=\frac{E_n(0)}{\pm\sqrt{E_n(0)^2+k^2}+k}$ and
\begin{equation*}
f_{n}(x)=\mbox{sech}^{-n+\omega}(\alpha x)
{}_2F_{1}\left(-n,1-n+2\omega,1-n+\omega,\frac{1}{1+e^{2\alpha x}}\right).
\end{equation*}
For $n=0$, the (non-normalized) bound states $\tilde F_0^\pm$ are $k$-independent. They are given by $\tilde{F}_0^+(x)=\left(\mbox{sech}^{\omega} (\alpha x),0\right)^T$ and $\tilde{F}_0^-(x)=\left(0,\cosh^{\omega} (\alpha x)\right)^T$, respectively, and  satisfy 
$$\tilde{H}_K(x,k)\tilde{F}_{0}^\pm(x)=\pm k \tilde F_0^\pm.$$

\begin{figure}
	\centering
	\begin{tabular}{cc} a) $c=0.15, b=0.1$, $v=0.432v_F$&b) $c=0.15, b=0.025$, $v=0.446v_F$\\
		\includegraphics[scale=.5]{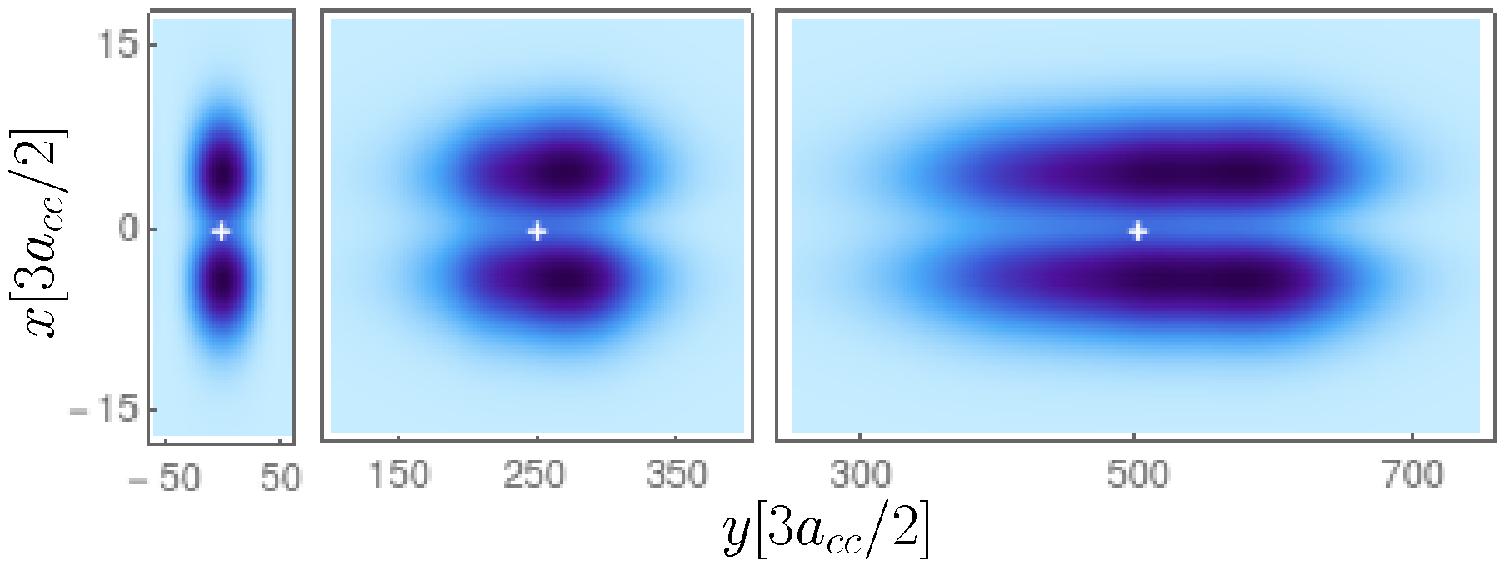}&\includegraphics[scale=.5]{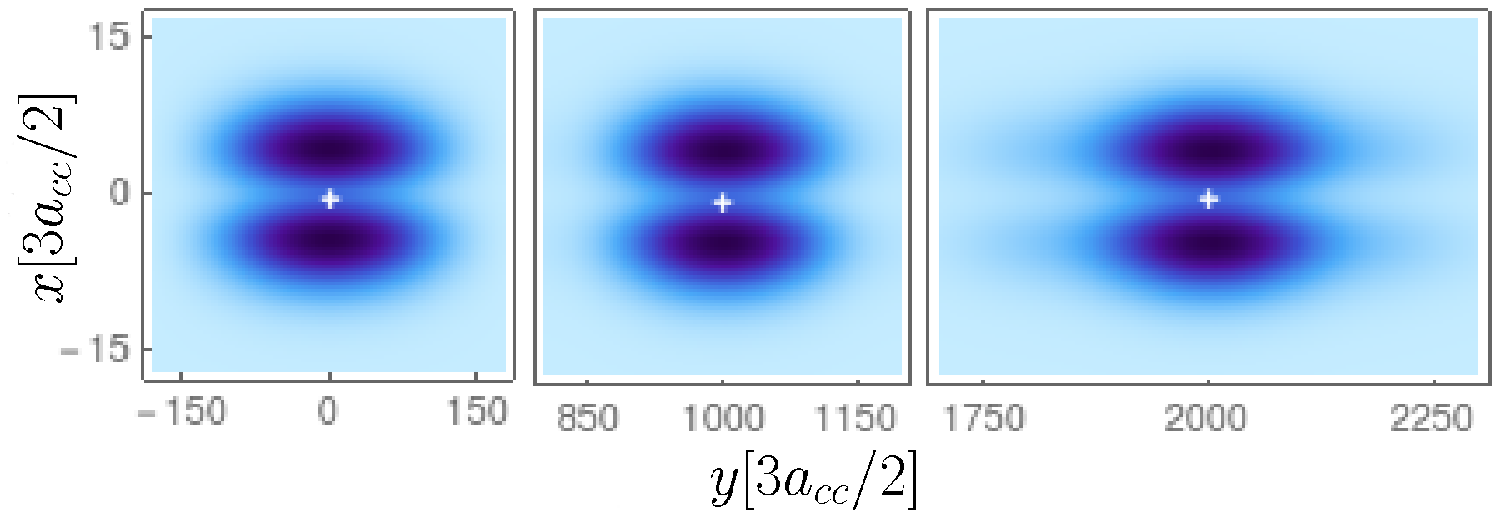}\\
		c) $c=0.215, b=0.018$, $v=0.582v_F$&d) Dispersionless packet, $c=0.15, b=0.1$\\
		\includegraphics[scale=.5]{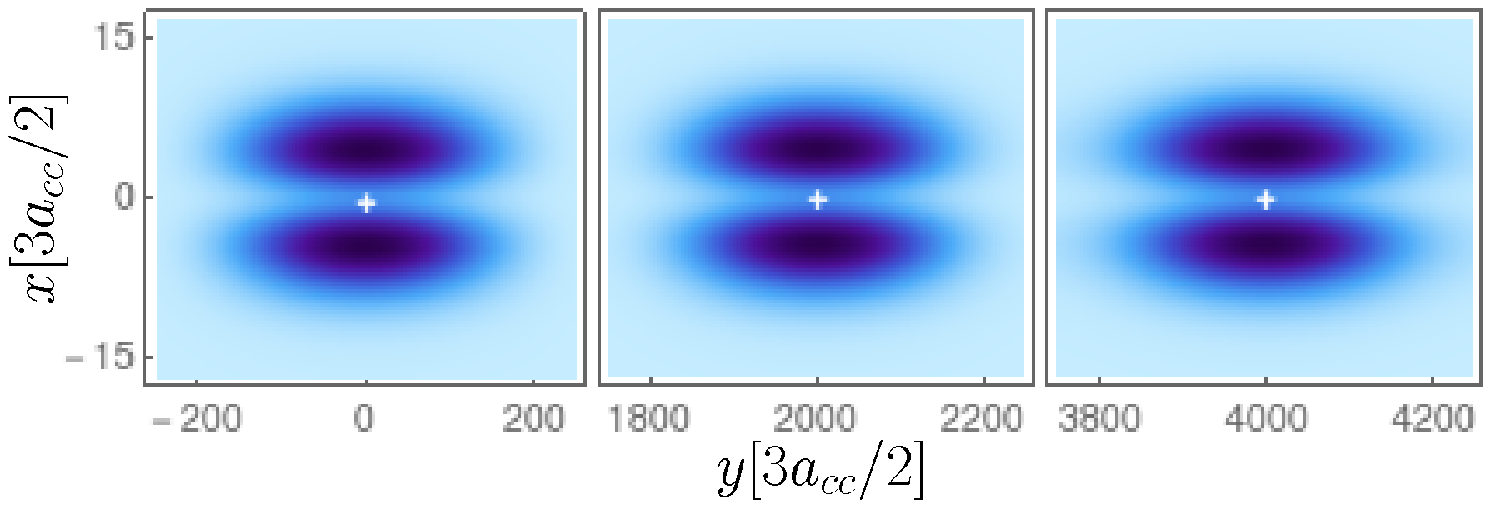}&\includegraphics[scale=.5]{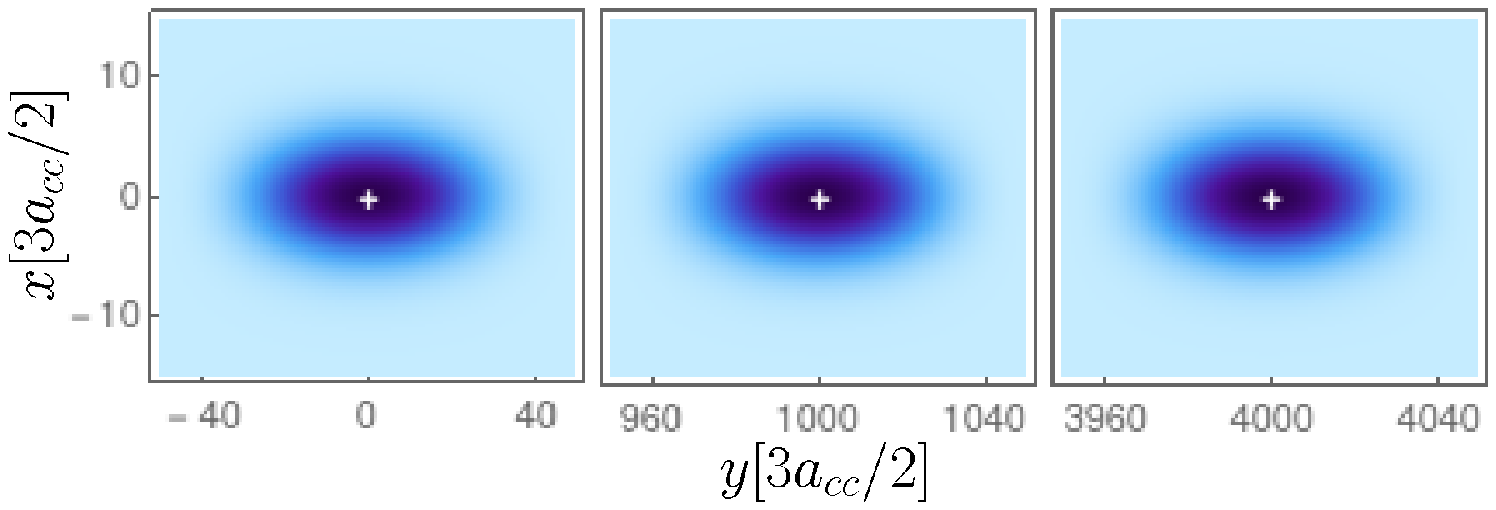}\\
		\multicolumn{2}{c}{e) $|A|^2=|A|^2\left(\frac{d}{v}\right)$}\\
		\multicolumn{2}{c}{\includegraphics[scale=.8]{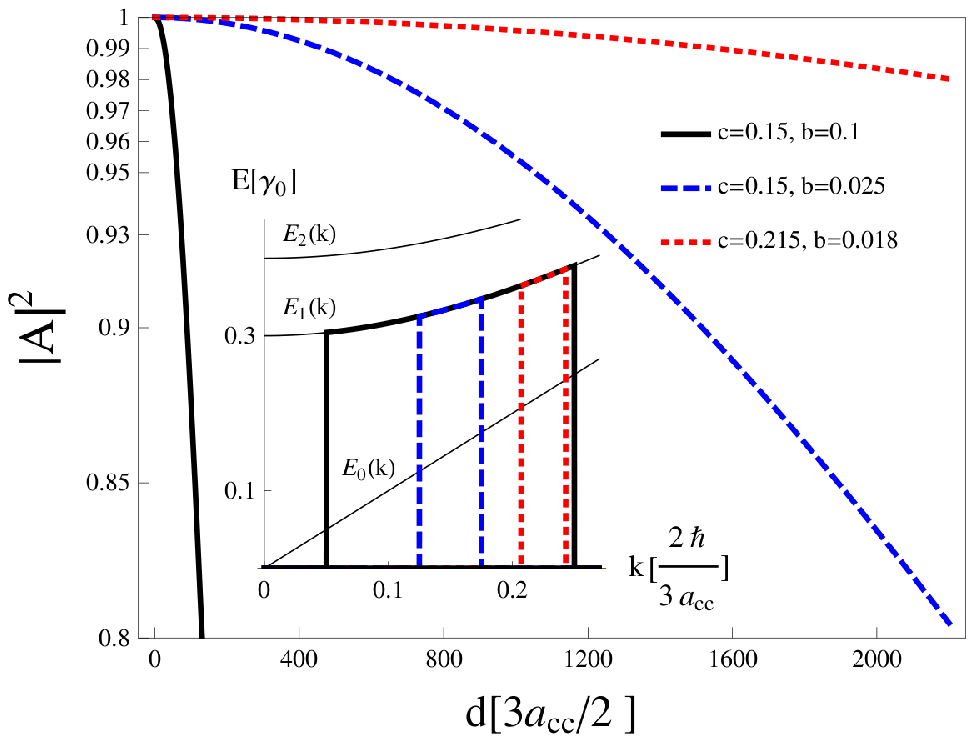}}
	\end{tabular}
	\caption{(Color online) a-c) Time evolution of $|\tilde\Psi_1|^2$ given by (\ref{eq:disp_pack}) with several specific choices of $c$ and $b$. d) Time evolution of $|\tilde\Psi_0|^2$ given by (\ref{eq:nondisp_pack}) with the same choice of the coefficient function as in a). a-d) In $m(x)$, $\omega=5$ and $\alpha=0.01$. The darker a color is, the higher value of the function is attained. The time evolution is compared with that of a classical free particle (represented  by a white cross) moving uniformly in the positive $y$-direction with the speed $v$ given by \eqref{eq:prop_vel}. e) Transition probability (\ref{A_def}) for the wave packets a)-c). The inset figure illustrates the first three energy bands together with the supports of the chosen coefficient functions. }
\label{fig:numerics}
\end{figure}

Let us consider the situation when the wave packets $\tilde \Psi_1$ and $\tilde \Psi_0$ are composed from $\tilde{F}_1^+(x,k)$ and $\tilde{F}_0^+(x)$ that correspond to the energy band $E_1(k)$ and $E_0(k)=k$, respectively. For definiteness, we take the following coefficient function
\begin{equation}\label{beta1}
 \beta_1(k)=
          C_{b} \exp{\left(-\frac{1}{b^2-(k-c)^2}\right)}
\end{equation}
fixed to $0$ outside $I_1=(c-b,c+b)$.
Here $b>0,\, c\in\R$, and $C_b$ is a numerical constant that ensures the normalization of
\begin{equation}\label{eq:disp_pack}
 \tilde \Psi_1(x,y)=\int_{I_1}e^{iky}\beta_1(k) \tilde{F}_1^+(x,k)\dd k
\end{equation}
and
\begin{equation}\label{eq:nondisp_pack}
 \tilde \Psi_0(x,y)=\tilde{F}_0^+(x)\int_{I_1}e^{iky}\beta_1(k)\dd k,
\end{equation}
respectively. The shape of the wave packets differ from the Gaussian ones, whose time evolution  in the graphene was studied e.g. in \cite{GWPG1}-\cite{GWPG3}.
In numerical calculations, we fixed the parameters so that $E_1(k)<E_2(0)$ for all $k\in I_1$. We computed the probability amplitude $A(t)$ as well as the density of probability of the wave packet in different stages of its evolution. The actual choice of $c$ and $b$ has profound impact on the dispersion of the wave packet, see Fig. \ref{fig:numerics} for illustration.

\section{Conclusion } 
We showed that the systems with a translational invariance  \eqref{eq:comm} and an energy band in its spectrum, see (\ref{H(x,k)E}), can host a normalizable wave packets that are dispersionless along the $x$-axis. When the energy band  is linear at least locally \eqref{eq:lin_ev}, the system can host wave packets with soliton-like behavior (\ref{nd1}). These do not disperse and move with a uniform speed.

The ideal situation described by (\ref{eq:lin_ev}) in its pristine form is unlikely to appear in an experiment. The situation was analyzed where the e\-nergy band ceased to be linear in $k$.  In that case, there can exist wave packets with very slow dispersion provided that the energy band is well approximated by a linear function on a finite interval.

We studied the situation where the perturbation $V=V(x,y)$  breaks down the translational invariance of the system (\ref{HV}). The soliton-like behavior of the wave packet is robust to the first order in the perturbation series \eqref{eq:asy_text}. In fact, this series lacks the linear term for any admissible state, not only for the dispersionless states, see Appendix. The second order term tends to zero as long as $V(x,y)$ is localized far enough from the channel where the wave packet propagates. If the perturbation is a translation invariant electrostatic potential with the symmetry axis perpendicular to the symmetry axis of the unperturbed system then the second order term takes very simple form given by \eqref{Dyson_1_spec} and \eqref{Dyson_2_spec}.

We focused on the systems described by the Dirac operator (\ref{hm}). 
However, our results are applicable to an ample class of systems  with the Hamiltonian that is decomposable into the direct integral and whose fiber operators $H(x,k)$ possess discrete eigenvalues \eqref{H(x,k)E}. 
In this context, the analysis of the bilayer graphene in the presence of topologically nontrivial electric field is worth mentioning. Energy bands, in the form of  mildly bent functions of $k$ were predicted theoretically \cite{BiGrapheneT} and, recently, the predictions were confirmed experimentally \cite{BiGrapheneE}. This system might be a promising candidate for observation of the wave packets described in this article.

Realization of the dispersionless (or slowly-dispersing) wave packets will be demanding on the precision of their preparation.  However, it was reported that fine tuned laser pulses can be utilized for creation and precise control of the wave packets orbiting the Rydberg atoms \cite{QC1}, \cite{QC2} or in coherent control experiments with the solid para-hydrogen \cite{QC3}. These results make the outlook towards experiments with dispersionless wave packets in the Dirac systems rather optimistic.

\section*{Acknowledgments}
\vspace{1mm}
VJ was supported by GA\v CR grant No.\ 15-07674Y.  MT was supported by GA\v CR grant No. 17-01706S.

%%%CHANGING THE COUNTER%%%%%%%
\setcounter{section}{1}
\renewcommand{\theequation}{\Alph{section}.\arabic{equation}}
%%%%%%%%%%%%%%%%%%%%%%%%%%%%%%

\section*{Appendix-Dyson expansion}
Let a total Hamiltonian has the following decomposition
$$H=H_0+V,$$
where $V$ is a bounded symmetric operator. The so-called Dyson expansion yields \cite[Example 9.5.5]{BlExHa}
\begin{equation*}
 \psi[t]=\ee^{-\frac{i}{\hbar}t H}\psi=\varphi_0[t]+\sum_{n=1}^{+\infty}(-i)^n\varphi_n[t],
\end{equation*}
where $\psi\equiv\psi[0]$ is a normalized state at $t=0$, $\psi[t]$ the state at time $t$, and 
\begin{align*}
 &\varphi_0[t]:=\ee^{-\frac{i}{\hbar}t H_0}\psi\\
 &\varphi_n[t]:=\int_0^t\dd t_1\int_0^{t_1}\dd t_2\ldots\int_0^{t_{n-1}}\dd t_n\ee^{-\frac{i}{\hbar}(t-t_1)H_0}V\ee^{-\frac{i}{\hbar}(t_1-t_2)H_0}V\ldots
 \ee^{-\frac{i}{\hbar}(t_{n-1}-t_n)H_0}V\ee^{-\frac{i}{\hbar}t_n H_0}\psi.
\end{align*}
Clearly, $\|\varphi_0[t]\|=1$ and for $n\in\N$,
\begin{equation*}
\|\varphi_n[t]\|\leq \frac{(\|V\|t)^n}{n!}, 
\end{equation*}
where $\|V\|$ stands for the operator norm of $V$.\footnote{If $V$ is the multiplication operator by a bounded real function $\mathcal{V}$, then $\|\mathcal{V}\|=\sup|\mathcal{V}(\vec{x})|$.}
If we truncate the expansion as follows
\begin{equation*}
 \psi_N[t]:=\varphi_0[t]+\sum_{n=1}^{N}(-i)^n\varphi_n[t]
\end{equation*}
then 
\begin{equation}\label{eq:Dyson_conv}
\|\psi[t]-\psi_N[t]\|\leq\sum_{n=N+1}^{+\infty}\frac{\|V\|^nt^n}{n!}\leq\frac{1}{(N+1)!}\sum_{N+1}^{+\infty}\|V\|^nt^n\leq \frac{1}{(N+1)!}\frac{(\|V\|t)^{N+1}}{1-\|V\|t}.
\end{equation}
Here we assumed that $\|V\|t<1$. Consequently, we may write
$$\psi[t]=\psi_N[t]+\mathcal{O}(t^{N+1})$$
as $t\to 0$. This approximation works for large times, too, as long as $\|V\|$ is relatively small, $\|V\|t\ll 1$.

Let us  compute the following transition amplitude
\begin{equation*}
 A(t):=\langle\ee^{-\frac{i}{\hbar}tH_0}\psi,\ee^{-\frac{i}{\hbar}tH}\psi\rangle=\langle\varphi_0[t],\psi[t]\rangle
\end{equation*}
by means of its approximations
\begin{equation}\label{eq:transition_A}
 A_N(t):=\langle\varphi_0[t],\psi_N[t]\rangle=1+\sum_{n=1}^{N}(-i)^n \alpha_n(t),
\end{equation}
where 
\begin{equation*}
 \alpha_n(t):=\int_0^t\dd t_1\int_0^{t_1}\dd t_2\ldots\int_0^{t_{n-1}}\dd t_n\langle\psi,\ee^{\frac{i}{\hbar}t_1H_0}V\ee^{-\frac{i}{\hbar}(t_1-t_2)H_0}V\ldots
 \ee^{-\frac{i}{\hbar}(t_{n-1}-t_n)H_0}V\ee^{-\frac{i}{\hbar}t_n H_0}\psi\rangle.
\end{equation*}
In particular, we have
\begin{equation*}
 \alpha_1(t)=\int_0^t\dd t_1\langle\ee^{-\frac{i}{\hbar}t_1H_0}\psi,V\ee^{-\frac{i}{\hbar}t_1H_0}\psi\rangle=\int_0^t\dd t_1\langle\varphi_0[t_1],V\varphi_0[t_1]\rangle
\end{equation*}
and
\begin{equation*}
 \alpha_2(t)=\int_0^t\dd t_1\int_0^{t_1}\dd t_2\langle\ee^{\frac{i}{\hbar}t_1H_0}V\varphi_0[t_1],\ee^{\frac{i}{\hbar}t_2H_0}V\varphi_0[t_2]\rangle
\end{equation*}

Remark that $\psi_N[t]$ in \eqref{eq:transition_A} is not normalized, so strictly speaking $A_N(t)$ is not a transition amplitude. However, it approximates $A(t)$ to the same order in $t$ as the normalized quantity $A_N(t)/\|\psi_N[t]\|$. Indeed, by the Cauchy-Schwarz inequality,
$ \alpha_n(t)\leq\|\varphi_n[t]\|$. Mimicking the estimates \eqref{eq:Dyson_conv} we arrive at 
\begin{equation*}
 |A(t)-A_N(t)|=\mathcal{O}(t^{N+1})
\end{equation*}
and 
\begin{multline*}
 \left|A(t)-\frac{A_N(t)}{\|\psi_N[t]\|}\right|=\left|A(t)-\frac{A_N(t)}{\|\psi[t]\|+\mathcal{O}(t^{N+1})}\right|=\left|A(t)-A_N(t)(1+\mathcal{O}(t^{N+1}))\right|\\
 \leq|A(t)-A_N(t)|+|A_N(t)|\mathcal{O}(t^{N+1})=\mathcal{O}(t^{N+1}). 
\end{multline*}

We will compute the approximation of the transition probability $|A(t)|^2$ up to the second order. We have
\begin{multline}\label{Aexp}
|A(t)|^2=(1-i\alpha_1(t)-\alpha_2(t)+\mathcal{O}(t^3))(1+i\overline{\alpha_1(t)}-\overline{\alpha_2(t)}+\mathcal{O}(t^3))\\
=1+2\Im\alpha_1(t)-2\Re\alpha_2(t)+|\alpha_1(t)|^2+\mathcal{O}(t^3).
\end{multline}
Therefore, it is sufficient to compute the terms that appear in $A_2(t)$, i.e., $\alpha_1(t)$ and $\alpha_2(t)$. Since $V$ is a symmetric operator, $\alpha_1(t)$ is real and $\Im\alpha_1(t)=0$.  Hence, the term linear in $t$ is absent in \eqref{Aexp}. If we introduce
\begin{equation*}
 \varphi[t]:=\ee^{\frac{i}{\hbar}tH_0}V\varphi_0[t]
\end{equation*}
and
\begin{equation*}
 \tilde\alpha_2(t):=\left\|\int_0^t\dd t_1\varphi[t_1]\right\|^2=\int_0^t\dd t_1\int_0^t\dd t_2\langle\varphi[t_1],\varphi[t_2]\rangle.
\end{equation*}
then
\begin{multline*}
 \tilde\alpha_2(t)=\int_0^t\dd t_1\int_0^{t_1}\dd t_2\langle\varphi[t_1],\varphi[t_2]\rangle+\int_0^t\dd t_1\int_{t_1}^t\dd t_2\langle\varphi[t_1],\varphi[t_2]\rangle\\
 =\alpha_2(t)+\int_0^t\dd t_2\int_{0}^{t_2}\dd t_1\overline{\langle\varphi[t_2],\varphi[t_1]\rangle}=\alpha_2(t)+\overline{\alpha_2(t)}=2\Re\alpha_2(t).
\end{multline*}
We arrive at
\begin{equation}\label{eq:asy}
 |A(t)|^2=1-\tilde\alpha_2(t)+\alpha_1(t)^2+\mathcal{O}(t^3).
\end{equation}
Remark that by the Cauchy--Schwarz inequality, 
\begin{equation*}
 \alpha_1(t)^2=|\alpha_1(t)|^2=|\langle\varphi_0[t],\varphi_1[t]\rangle|^2=|\langle\psi,\int_{0}^{t}\dd t_1\varphi[t_1]\rangle|^2\leq \left\|\int_0^t\dd t_1\varphi[t_1]\right\|^2=\tilde\alpha_2(t).\\
\end{equation*}
Therefore the correction in \eqref{eq:asy} is always non-positive.

\end{document}